# Smart Novel Computer-based Analytical Tool for Image Forgery Authentication

Rozita Teymourzadeh,CEng. IEEE, IET Member, Amirrize Alpha Laadi M., VH MOK, IET Member
Faculty of Engineering, Technology & Built Environment
UCSI University, 56000, Malaysia
rozita@ucsi.edu.my

*Abstract*— **This paper presents an integration of image forgery detection with image facial recognition using black propagation neural network (BPNN). We observed that facial image recognition by itself will always give a matching output or closest possible output image for every input image irrespective of the authenticity or otherwise not of the testing input image. Based on this, we are proposing the combination of the blind but powerful automation image forgery detection for entire input images for the BPNN recognition program. Hence, an input image must first be authenticated before being fed into the recognition program. Thus, as image security identification and authentication requirement, any image that fails the authentication/verification stage is not to be used as an input/test image. In addition, the universal smart GUI tool is proposed and designed to perform image forgery detection with the high accuracy of ±2% error rate. Meanwhile, a novel structure that provides efficient automatic image forgery detection for all input test images for the BPNN recognition is presented.**

## I. INTRODUCTION

Face pattern recognition is an exciting area of research as a result of scientific issue and possible social, legal and security application [1]. Two approaches to 2-D features identification are base on position and based on geometric approach [2].

In 1987, Sirovich and Kirby [3] illustrated in their approach that a particular facial images reasonably represents in terms of the maximum coordinates branded as Eigen image. It was observed that an acceptable picture of a face could be reconstructed from the specification of gray levels at 214 pixel locations. However, they showed that in actual construction, only 40 numbers with admixtures of Eigen pictures can be characterized a face.

Wu and Haung [4] were used 24 measurements in fiducially brands profile double facial image. The proposed system has been reduced by recognizing performance when more image in the database because there are not enough to differentiate features that identify a user.

Then, Turk and Penland [5] introduced the Eigen faces method that achieved a remarkable acknowledgement speed, but the speed drastically reduced, when against resizing. Wiskott [6] took advantage of scale information to stage the information obtained from Gaber wavelet transform the facial images.

Later on, in 2006, Ahonen and Hadid [7] commenced the local binary pattern (LBP). This operator is for measuring the structure information for LAN of gray scale image. Because of direct interactions, indifference to the lights, ability to retrieve picture detail, the user can get the model own region which are more favorable.

Gallagher [8] analyzed a set of images and proposed that image forgery most often includes some form of geometric transformation.

The various research work had been completed [9-13,16] to reduce the efficiency with variations in illumination, image size, size of database, percentage error. In addition, inability to detect altered images for more realistic recognition still is challenging task.

Hence, this paper proposes an efficient passive evaluation tool for authenticity of input images before recognition process stage, furthermore, the adjustment of illumination of images using contrast stretching and image histogram equalization techniques will be performed. In this proposed research work, the quality percentage is adjustable in order to achieve high performance (as low as 2% error rate) with a dual database approach. The test image is reduced to a lower 1-D dimensional vector, which represents distinguishing features of the test image.

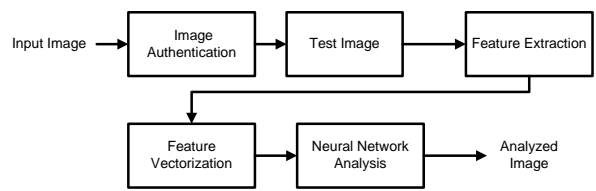

Figure 1. Proposed process structure

The 1-D image vectors are fed into the neural network, the Euclidean distance of the 1-D image vector is compared to each test database, and the closest match is found and outputted.

## II. SYSTEM DESIGN FUNDAMENTAL

The focus of the image authentication stage (Fig. 1) is the presence of the image re-sampling trace and/or interpolation signal. A derivative operator on the variance of the image will detect these traces. Later, the signal is processed through radon transformation that results in a periodic signal embedded or completely imposed on the image spectrum graph [8]. Hence, interpolated signal and re-sampled signal will be ready to function.

**A. Interpolated Signal:** There are two main stages in geometric transform [14,15]. In the first stage, a spatial transformation of the physical re-arrangement of pixels in the image is performed and can be represented by a transformation function T.

$$X' = T_X(X,Y) \quad Y' = T_Y(X,Y) \tag{1}$$

As earlier stated, the geometric operations that are common with most forgeries are re-scaling, rotation and skewing. Hence, the importance of detection will be traced wit utilization of affine transformations.

The general equations for affine transformations are given by:-

$$X' = a_0 + a_1 x + a_2 y$$
$$Y' = b_0 + b_1 x + b_2 y \tag{2}$$

The second stage is the interpolation. The image brightness of the processed image is allocated using low pass interpolation filter. If the Nyquist condition is met, the range of $F(w)$ do not overlap in the Fourier domain the original signal $f(x)$ can be rebuilt entirely[12].

Combining the derivative theorem with the convolution theorem leads to the conclusion that by convolution of $f_k$ with derivative kernel $D_n(w)$, it is possible to reconstruct the nth derivative of the image $f(x)$. The result of interpolated operation of $f_w(x)$ is denoted by;

$$f^w(x) = \sum_{k=-\infty}^{\infty} f_k w \left( \frac{x}{\Delta x} - k \right)$$

$$D^n\{f^w\}(x) = D^n \left\{ \sum_{k=-\infty}^{\infty} f_k w \left( \frac{x}{\Delta x} - k \right) \right\} =$$

$$\sum_{k=-\infty}^{\infty} f_k D^n \{w\} \left( \frac{x}{\Delta x} - k \right) \tag{3}$$

By assuming the constant variance random process, then the variance of $D^n\{f^w\}$ which is var $\{D_n\{f_w\}(x)\}$ as a function of $x$ is given by;

$$var\{D^n\{f^w\}(x)\} = R_{D^n}\{f^w\}(x,x) =$$
$$\sigma^2 \sum_{k=-\infty}^{\infty} D^n\{w\} \left( \frac{x}{\Delta x} - k \right)^2 \tag{4}$$

Similarly, the covariance is represented as;

$$R_{D^n}\{f^w\}(x, x+\xi) = \sigma^2 \sum_{k=-\infty}^{\infty} D^n\{w\} \left( \frac{x}{\Delta x} - k_1 \right)$$
$$x D^n\{w\} \left( \frac{x+\xi}{\Delta x} - k_2 \right) \tag{5}$$

Now, by assuming that $\vartheta$ is an integer, it can be noticed that;

$$var\{D^n\{f^w\}(x)\} = var\{D^n\{f^w\}(x + \vartheta \Delta)\} \tag{6}$$

Thus, $var\{D_n\{f_w\}(x)\}$ is periodic over $x$ with period $\Delta x$ that $\Delta x$ is the sampling step.

It is verified the periodicity by the following;

$$var\{D^n\{f^w\}(x + \vartheta \Delta)\}$$

$$= \sigma^2 \sum_{k=-\infty}^{\infty} D^n\{w\} \left( \frac{x+\vartheta \Delta_x}{\Delta x} - k \right)^2$$

$$= \sigma^2 \sum_{k=-\infty}^{\infty} D^n\{w\} \left( \frac{x}{\Delta x} - (k - \vartheta) \right)^2 = var\{D^n\{f^w\}(x)\} \tag{7}$$

**B. Re-Sampled Signal:** From equation (4), it is clear that different interpolators will change the structure of the original signal in different ways. The resulting periodic variance function computed using equation (4) for the nearest-neighbor interpolation. Hence, marks inserted in this interpolating are easy to spot using derived operator for them [12]. It is continuous but its first derivative is discontinuous. Cubic interpolate using a third-order extrapolation polynomial kernel. Then, the $n^{th}$ derivative of the investigated 2 signal $b(x,y)$, $D_n\{b(x,y)\}$ is computed.

**C. Radon Transformation:** Radon transform is applied for traces of affine transformation. The Radon transformation computes projections of magnitude of $D_n\{b(x,y)\}$ along specified direction determination by an angle $\theta$.

The projection is a line integral in a certain direction. This line integral is expressed as;

$$\rho D^n\{b\}(x,y) = \int_L |D^n\{b(x,y)\}| \, dl \tag{8}$$

By assuminy that;

$$\begin{bmatrix} x' \\ y' \end{bmatrix} = \begin{bmatrix} cos\theta & sin\theta \\ -sin\theta & cos\theta \end{bmatrix} \begin{bmatrix} x \\ y \end{bmatrix} \tag{9}$$

It is possible to represent the Radon transform in the following way;

$$\rho\theta(x') = \int_{-\infty}^{\infty} D^n\{b(x,y)\} \cdot (x' cos\theta - y' sin\theta, x' sin\theta + y' cos\theta) dy' \tag{10}$$

To compute the Radon transformation, pixels are divided into four sub pixels and each sub pixel is projected separately. The Radon transformation is computed at angles $\theta$ from 0 to 179 degrees, in 1 degree increments. Hence, the output of this is 180 1-D vectors, $\delta\theta$ ($\theta$ is the orientation of the $X'$ axis counterclockwise from the $x$-axis). the corresponding auto covariance sequences of $\delta\theta$ contain a specific strong periodicity, if the investigated signal has been re-sampled. The auto covariance sequences of $\rho\theta$ is computed as;

$R_{\rho\theta}(k) = \sum_i (\rho\theta\,(\,i+k\,) - \overline{\rho\theta})(\rho\theta\,(\,i\,) - \overline{\rho\theta})$
(8)

The project target is to determine the image being investigated has undergone affine transformation. Hence, we focus only on the strongest periodic patterns present in the auto covariance $R\,\delta\theta$.

**D. Image Recognition:** It is difficult to recognize the face with ranking issue and classification process. Hence, the proposed technique used artificial nerve cells for training classifier. In recent years, BPNN system [1] makes use of fractal encoding method. It presented as input to the BPNN for identification purposes. However, in order to complete image authentication and recognition in the research work, four stages are introduced.

### III. PROPOSED SYSTEM STRUCTURE

The proposed research work deals with an integration of four phases. The initial phase deals with image verification (originality authentication) with neural network pre-processing training and testing phases. Fig. 2 shows the flowchart performance of this research work that will be discussed in four stages.

**A. Image verification (Phase 1):** The image verification phase is performed using detecting traces of image tampering such as re-sampling or interpolation or both. In this phase, a test/input process involves derivative operators and radon transformation. This process produces a periodic pattern in the image spectrum if the test image is forged or it produces simple impulse signals if the test image is still original. Meanwhile, It is observed that both image spectra forms could be overlapped that produces a form of image tampering. However, the proposed system specification allows selecting only original test or input images.

**B. Image-preprocessing (Phase 2):** The image preprocessing as second phase is ready to function. In the pre-processing phase, time effective preprocessing is performed in order to make image data best fit for neural network input. Average filtering is applied and contrast of the image is enhanced through histogram equalization process.

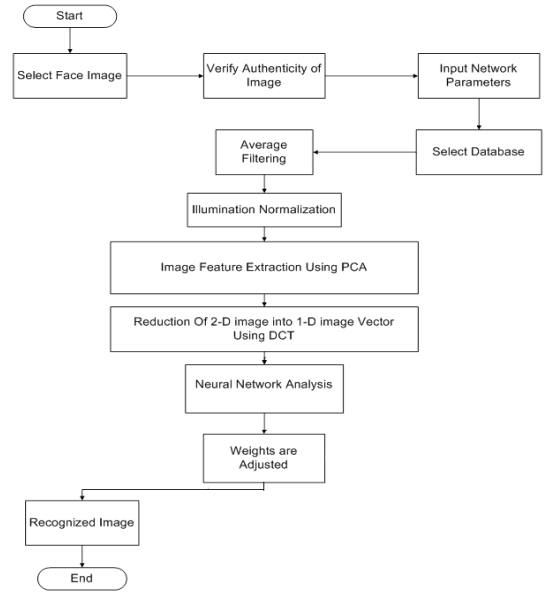

Figure 2. Proposed system performance flowchart

Then the size of image will be reduced so as to make it small but effective and most suitable for neuronal processing stage. Before moving images into the neural processing phase, all images data (test image and training database images) undergo a process of vectorization that is conversion of 2-D images into 1-D vectors. This conversion is due to neural network requires 1-D vectors for processing and conversion is performed using principal component analysis (PCA) and 2-D discrete cosine transform. Figure 3 shows the vectorization of one original image.

**C. Neural Network Training (Phase 3):** The third phase is the neural network training. The structure of the proposed neural network is based upon multi-layers that are input, hidden and output layer. Pre-processed images in the vector format are presented as input to the neural network, which contains 400 neurons in the input layer and 90 neurons in the hidden layer. The number of neurons in the hidden layer is determined by experiences and guesswork considering optimal performance.

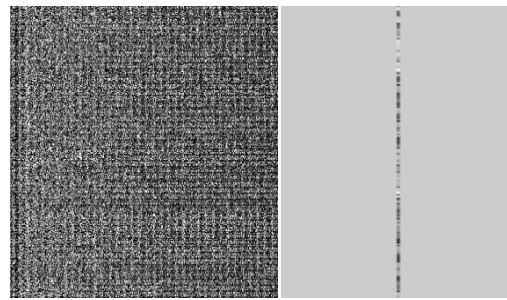

Figure 3. Image vectorization

As a final point, production level contains nerve cells that are the number of objects is being considered. The algorithm for effective use of a neural networks and reduces the slope of errors through changing weight and offset continued with impetus. The neural network is trained upon some set of images and tested upon different set of images.

In the proposed method, neural network uses BPNN algorithm for error computation and new weight calculation for each neuron link. It returns the output of each level, extract the mean square error (MSE) and spread it back if it is not close to the target. The response of the neural network is reliant upon weights, biases and transfer functions.

The transfer functions make use of in the feed forward BPNN in intermediate, input and output layer. Fig. 4 shows the training graph of neural network while the system will find the original image.

**D. Testing stage (Phase 4):** The fourth phase is the testing of the neural network. Images for testing are applied to the trained neural network along with the already trained database images for calculating the percentage accuracy and error.

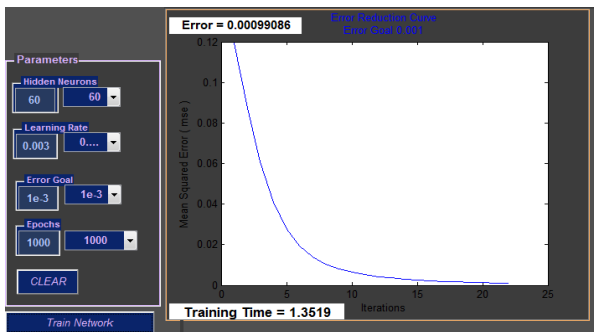

Figure 4. Neural network training curve

In testing phase, just like in the training phase, incoming images undergo all the pre-processing stage and are made available to the network for evaluation.

This test images (extracted face image) is then processed using the neural network analytic tool. After a number of iterations by the network through each image in the database, the error reduces based on the gradient descent-learning rule, the set error goal is reached and a matching image to test image is found as shown in the results session.

## IV. DISCUSSION

The two database containing training images is made up of ten images numbered (1-10) each, with the same size, format and dimension. The test images are sampled and cropped with the different position. Fig. 5 and 6 show the image spectrum respectively.

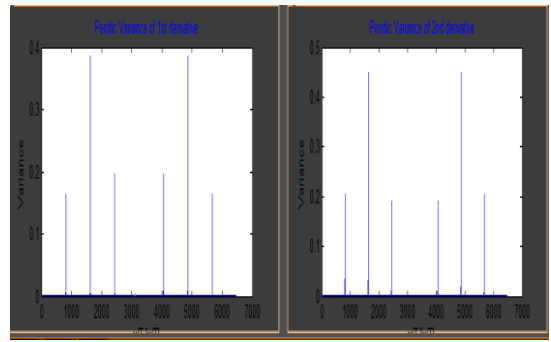

Figure 8. Original image spectrum

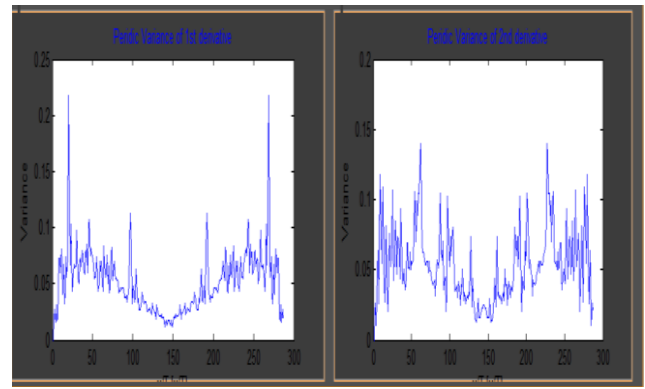

Figure 6. Forged image spectrum.

These spectra are resulted from original and forged images respectively. The difference in spectrum is seen as a kind of sinusoidal periodic pattern, which replaces or overlaps the original impulse signals or spikes of an original image. These are the result of obtaining the first and second derivative of the auto covariance of the images. The auto covariance is obtained using the radon transformation through 0 to 179 degrees. Hence, it is a proved for efficiency of proposed system to detect the authenticity of images. The major integral part of the proposed analytical tool is the capability to extract a face from a full image and run it through any given database for a matching or equivalent face image. Fig. 7 illustrates the recognition of input image while different position is applied.

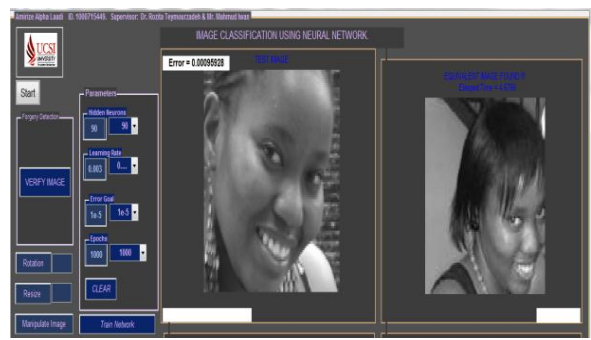

Figure 7. Image recognition

The high-speed performance is determined by the proper selection of the number of hidden neurons. The larger amount of hidden neurons results the faster network converges. Fig. 8 shows the time consuming required when the hidden layers in neural network processor are increased.

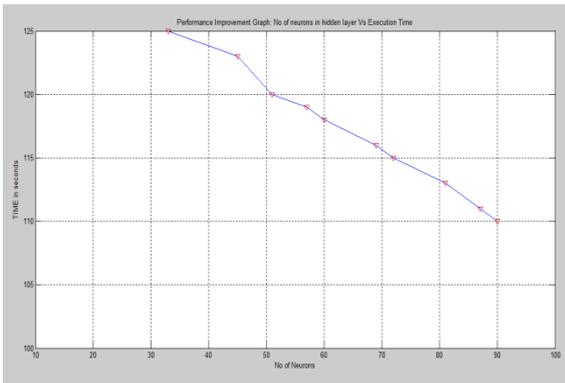

Figure 8.  Neurons in hidden layer vs. execution time

Fig. 9 shows the error percentages of proposed analytical tools when the number of imaged are increased. As shown in Fig. 9, it is found that with increasing the number of subjects, performance error that is percentage of recognition with utilizing PCA and BPNN approach will decrease slightly.

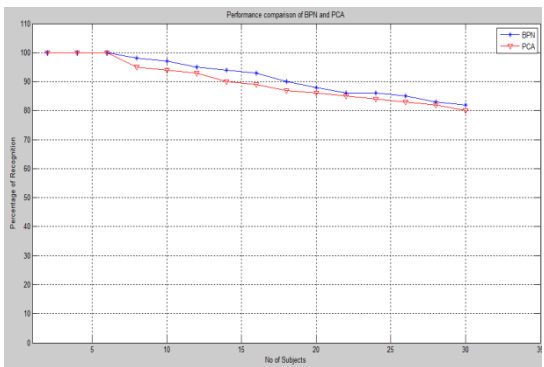

Figure 9.  Recognition percentage vs. number of subjects

## V.  CONCLUSION

Face recognition as a classification problem does not take into consideration the originality of the test image and thus can always output a matching image for even a forged input image. It is therefore, important to first verify the authenticity of any test image. A single smart BPNN image processing for face image recognition is only highly accurate with small number of subjects (images) and requires a lowering of the image resolution for a whole recognition task as a way of reducing the computational complexity.  Further work, can be done by further or better integration of both the image authentication phase with the rest of the neural network process such that we can eliminate the need for a human choice whether or not to proceed with the recognition of an input image after verification. Thus if an input image is found to have been forged or tampered with, the system automatically  rejects it and asks for an original of the rejected image before proceeding with the recognition task. A database of unlimited sizes, format and dimensions of train images is created and used as a source of target input for the neural network. The project was designed and investigated and it was found that the resolution improved by ±2% error rate when number of database images applied is less than 20 images.

REFERENCES

[1] Vinay Kumar B., Shreyas, B.S. and Ganesh Murthy C.N.S. 2007. A Back Propagation Based Face Recognition Model, Using 2D Symmetric Gabor Features. *IEEE Proc. of the signal processing, communications and networking*, pp. 433-437. DOI: 10.1109/ICSCN.2007.350776

[2]  Bolme D., Beveridge R., Teixeira M. & Draper B., 2003. The CSU Face Identification Evaluation System: Its Purpose, Features and Structure. *International Conference on Vision Systems*, pp 304-311, Graz, Austria, April 1-3. Published by Springer-Verlag. DOI: 10.1.1.89.1918

[3] Sirovich L. & Kirby M., 1987.  A Low Dimensional Procedure for the Characterization of  Human Faces. *Journal of  Optical Society of  American* , vol.4(3):519-524. DOI: 10.1364/JOSAA.4.000519

[4]  Wu C.J. & Huang J.S. 1990. Human face  profile recognition by computer. *Journal of Science Direct Pattern Recognition* . vo1. 23(3): 255-259. DOI: 10.1016/0031-3203(90)90013-B

[5] Turk M.A. & Pentland A. P. 1991. Face recognition using eigenfaces. *IEEE Conference on Computer Vision and Pattern Recognition,* pp. 586-591. DOI: 10.1109/CVPR.1991.139758

[6] Wiskott L., Fellous, J.M.,  Kuiger, N. and von der Malsburg, C. 1997. Face Recognition by Elastic Bunch Graph Matching. *IEEE Trans. On pattern Analysis and machine intelligence*,vol. 19(7):775- 779. DOI: 10.1109/34.598235

[7] Ahonen T., & Hadid A. 2006. Face recognition with Local Binary Patterns: Application on Face Recognition, *Proc. IEEE Transactions on Pattern Analysis and Machine Intelligence IEEE Press*, vol 28(12):2037-2041.DOI: 10.1109/TPAMI.2006.244

[8] Gallagher A. C. 2005. Detection of linear and cubic interpolation in JPEG compressed images,  *in Proc. IEEE Computer. Soc 2nd Canadian Conf. Computer*

*Robot Vision, Washington, DC*, pp. 65–72. DOI: 10.1109/CRV.2005.33

[9] Ming Hu, Qiang Zhang, Zhiping Wang, 2008. Application of Rough Sets to Image Pre-Processing for Face Detection. *IEEE Proc. Of the information and automation,* pp. 545-548. DOI: 10.1109/ICINFA.2008.4608060

[10] Mehryar S., Martin K., Plataniotis K.N. and Stergiopoulos S., 2010. Automatic landmark detection for 3D face image processing. IEEE Conference on Evolutionary Computation (CEC), pp.1-7. DOI:10.1109/CEC.2010.5586520

[11] Jaeyoung Kim and Heesung Jun, 2011. Implementation of image processing and augmented reality programs for smart mobile device. *IEEE Conference on Strategic Technology (IFOST). vol 2, pp.* 1070 – 1073. DOI:10.1109/IFOST.2011.6021205.

[12] Hou H. & Andrews H. 1978. Cubic splines for image interpolation and digital filtering, *IEEE Trans. Acoust., Speech Signal Process.*, vol. 26(6):508–517. DOI: 10.1109/TASSP.1978.1163154

[13] Heechul Han, Jingu Jeong and Arai, E. 2011. Virtual out of focus with single image to enhance 3D perception. *IEEE Conference on 3DTV The True Vision - Capture, Transmission and Display of 3D Video,* pp. 1 - 4 DOI: 10.1109/3DTV.2011.5877188

[14] Popescu A. C. & Farid H. 2005. Exposing digital forgeries by detecting traces of re-sampling, *IEEE Trans. Signal Process.*, vol. 53(2):758–767. DOI: 10.1109/ICOSP.2006.345714

[15] Prasad S. & Ramakrishnan K. R. 2006. On re-sampling detection and its application to image tampering, in Proc. *IEEE Int. Conf. Multimedia Expo., Toronto, ON, Canada*, pp.1325–1328. DOI: 10.1109/ICIEA.2009 5138406

[16] Lu J., Plataniotis K.N., & Venetsanopoulos A.N. 2003. Regularized Discriminate Analysis For the Small Sample Size Problem in Face recognition, *Science Direct Pattern Recognition Letters*, Science Direct vol. 24(16): 3079—3087. DOI: 10.1016/S0167-8655(03)00167-3